\documentclass[prb, twocolumn, superscriptaddress]{revtex4-1}
\usepackage{amsmath,amssymb,bm}
\usepackage{hyperref}
\usepackage{pdfpages}
\usepackage{graphicx}
\usepackage{epstopdf}
\usepackage{latexsym}
\usepackage{natbib}
\usepackage{braket}
\usepackage{float}
\usepackage[normalem]{ulem}
\usepackage{comment}
\usepackage{mathtools}
\usepackage{array}
\usepackage{tabu}
\usepackage{multirow}
\usepackage[inline]{enumitem}

\makeatletter
\AtBeginDocument{\let\LS@rot\@undefined}
\makeatother

\begin{document}
\title{Dc Electrical Current Generated by Upstream Neutral Modes}
\author{Ankur Das}
\email{ankur.das@weizmann.ac.il}
\affiliation{Department of Condensed Matter Physics, Weizmann Institute of Science, Rehovot, 76100 Israel}
\author{Sumathi Rao}
\affiliation{Harish-Chandra Research Institute, HBNI, Chhatnag Road, Jhunsi, Allahabad 211 019, India}
\affiliation{International Centre for Theoretical Sciences (ICTS-TIFR),
Shivakote, Hesaraghatta Hobli, Bangalore 560089, India}
\author{Yuval Gefen}
\affiliation{Department of Condensed Matter Physics, Weizmann Institute of Science, Rehovot, 76100 Israel}
\author{Ganpathy Murthy}
\affiliation{Department of Physics and Astronomy, University of Kentucky, Lexington, KY 40506, USA}

\newcommand\redsout{\bgroup\markoverwith{\textcolor{red}{\rule[0.5ex]{2pt}{0.4pt}}}\ULon}
\newcommand\bluesout{\bgroup\markoverwith{\textcolor{blue}{\rule[0.5ex]{2pt}{0.4pt}}}\ULon}

\newcommand\Oout{\bgroup\markoverwith{\textcolor{Orange}{\rule[0.5ex]{2pt}
{1.pt}}}\ULon}
\newcommand\blueout{\bgroup\markoverwith{\textcolor{Blue}{\rule[0.5ex]{2pt}
{1.pt}}}\ULon}
\newcommand\Rout{\bgroup\markoverwith{\textcolor{red}{\rule[0.5ex]{2pt}
{1.pt}}}\ULon}
\newcommand\Gout{\bgroup\markoverwith{\textcolor{green}{\rule[0.5ex]{2pt}
{1.pt}}}\ULon}

\newcommand{\AD}[1]{{\textcolor{blue}{\bf AD: #1}}}
\newcommand{\ADhide}[1]{{}}
\newcommand{\ADEDITED}[2]{{\bluesout{#1}}{\bf ~\textcolor{blue}{AD: #2}}}
\newcommand{\ADEDITOKAY}[2]{{}{\textcolor{black}{#2}}}
\newcommand{\GM}[1]{{\textcolor{red}{\bf GM: #1}}}
\newcommand{\GMEDITED}[2]{{\redsout{#1}}{\bf ~\textcolor{red}{GM: #2}}}
\newcommand{\YG}[1]{{\textcolor{violet}{\bf YG: #1}}}
\newcommand{\SR}[1]{{\textcolor{green}{\bf SR: #1}}}
\newcommand{\SREDITED}[2]{{\Gout{#1}}{\bf ~\textcolor{green}{SR:#2}}}

\begin{abstract}

Quantum Hall phases are gapped in the bulk but support chiral edge
modes, both charged and neutral. Here we consider a circuit where the
path from the source of electric current to the drain necessarily passes
through a segment consisting solely of neutral modes. We
find that upon biasing the source, a dc electric current is detected at the
drain, provided there is  backscattering between counter-propagating modes
under the contacts placed in certain locations. Thus, neutral modes carry
information that can be used to nonlocally reconstruct a dc charge current.
Our protocol can be used to detect any neutral mode that counterpropagates
with respect to all charge modes. Our protocol applies not only to the edge
modes of a quantum Hall system, but also to systems that have neutral modes
of  non-quantum Hall origin. We conclude with a possible experimental
realization of this phenomenon.

\end{abstract}

\maketitle
% \AD{Ankur}

% \SR{Sumathi}

% \YG{Yuval}

% \GM{Ganpathy}

% \ADEDITED{test}{}
% \SREDITED{test}{}
% \YGEDITED{test}{}
% \GMEDITED{test}{}

\section{Introduction}
The quantum Hall effects (QHE) \cite{Klitzing1980} are the earliest known example
of topological insulators \cite{Hasan2010}. They have a charge gap in the bulk, and
all currents are carried by edge/surface modes, which can be either charged (with
fractional charge in the fractional QHE) or neutral chiral modes. While the charge
modes produce quantized electrical conductance, neutral modes are a manifestation of
topology, electron-electron interactions, and possibly disorder, and contribute to
heat transport. Neutral edge modes in quantum Hall systems have been detected by shot
noise experiments \cite{Bid2010} and also by their quantized heat transport
coefficients \cite{Venkatachalam2012,Deviatov_etal_2011}.  Apart from
quantum Hall systems, neutral (e.g. Goldstone) modes arise in systems
in which a continuous symmetry is broken spontaneously.

In this work, we design a geometry where the unique current path from
the source to the drain is forced to pass through a segment consisting
of neutral modes only. We assume that the $U(1)$ symmetry of each channel
is broken by the contacts; thus backscattering between channels is
{present
under them}. The breaking of these $U(1)$ symmetries results in a non-zero dc current at
the drain D. This protocol can be used   either as a transformer, which
converts charge current to neutral current, and then back to charge current, or as an efficient detector of
neutral modes as long as the neutral counterpropagates with respect to all charge modes.

\begin{figure}[!ht]
\centering
\includegraphics[width=\columnwidth]{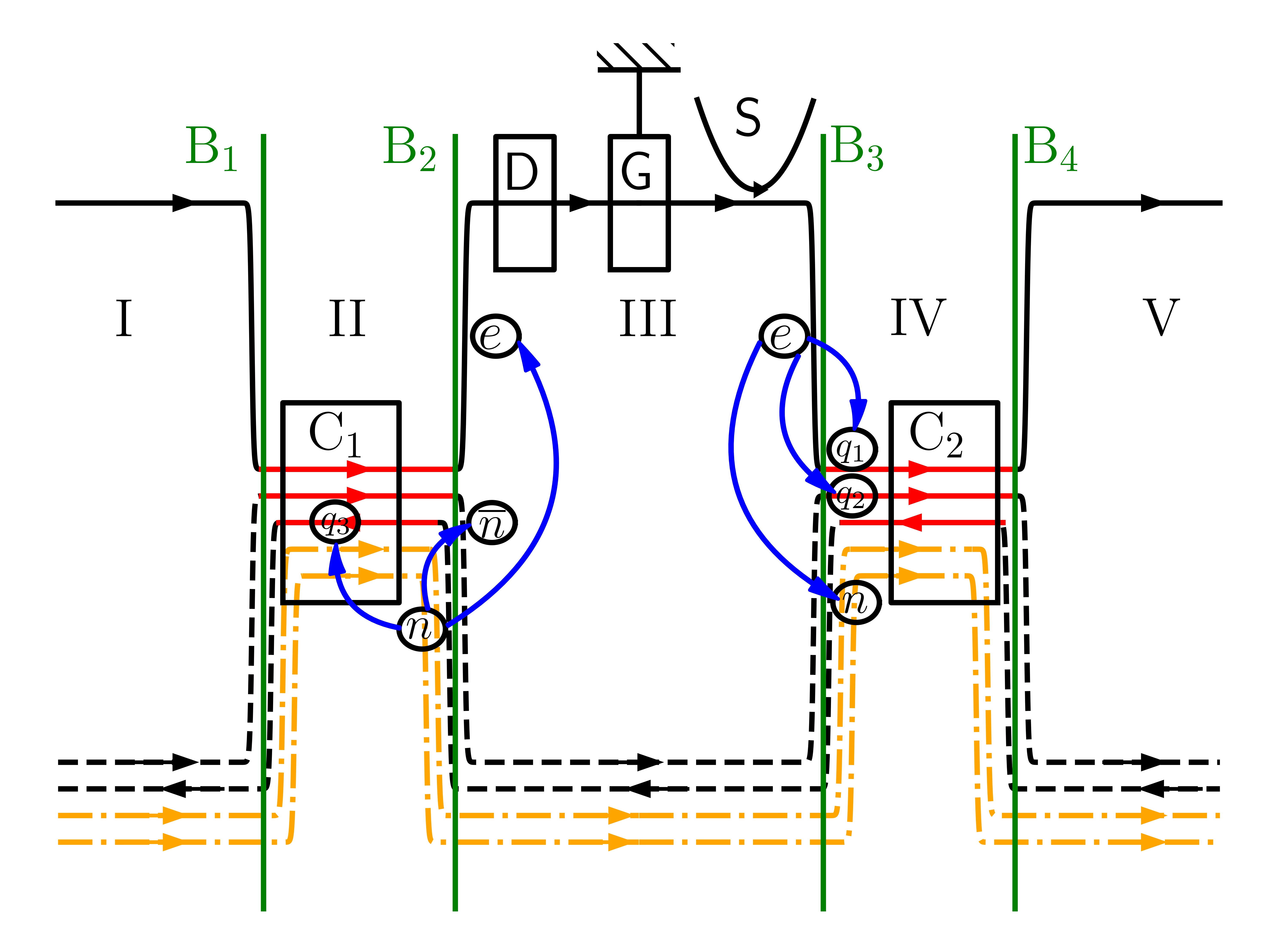}
\caption{A single right-moving chiral charged mode (solid black
line) represents the edge of a $\nu=1$ quantum Hall system
extending above the figure, which is the ``probe'' system. Charges
are injected at the source S and detected at the drain D. The
``test'' system extends below the figure, and has two
counterpropagating neutral modes (dashed black lines), and
possibly other charged chiral modes (dash-dotted
orange lines), which all have to be right-moving for our scheme to
be relevant. The edges of the two systems overlap only
in regions II and IV, separated from the active region III by
boundaries B$_2$, B$_3$. Density-density interactions between the
chiral modes of the top and bottom systems exist only in regions
II and IV, which also host the contacts C$_1$ and
C$_2$. Regions I and V are present to specify boundary
conditions. Tunneling/scattering between the chiral modes occurs solely under the
contacts. Reflection and transmission of a right-moving charge injected at S is
shown schematically at B$_3$, while a similar
process for a left-moving neutral excitation is shown at B$_2$.}
\label{fig:geo}
\end{figure}

The proposed geometry is shown in Fig. \ref{fig:geo}. The relevant physics
can be extracted by focusing on regions II, III, and IV. The solid black
line at the top is a right-moving chiral charge mode, arising from a
$\nu=1$ quantum Hall system extending above Fig. \ref{fig:geo},
constituting the ``probe'' system. The dashed lines at the bottom in region
III are a pair of counter-propagating gapless, bosonic \cite{bosonic},
neutral modes, presumed to arise from a ``test'' system extending below Fig.
\ref{fig:geo}.  The test system may be quantum Hall, so long as all its charge
modes (dash-dotted orange lines) are right-moving, or it may be a system with
neutral modes only, such as an $XXZ$ chain. Electrons are injected from the
source S via tunnelling into the probe chiral edge mode and detected at the
drain D. The source and drain are separated by a grounded contact G. Clearly,
current cannot flow from S to D along the right-moving, chiral top edge. The
edge modes of the probe and test systems overlap, and thus interact, only in
regions II and IV. The interaction is of the density-density form, with separate
number conservation in the ``bare'' charged and neutral modes. These
interactions renormalize the bare charged and neutral modes such that,
generically, all three renormalized eigenmodes have nonuniversal charge. Regions
II and IV also host the  contacts C$_1$ and C$_2$ respectively
\cite{Spanslatt_etal_2021}, which we assume can be decoupled from the
interacting modes at will. Finally, regions I and V are semi-infinite ``free''
regions, where the edges of the probe and test systems are fully decoupled and
are present to fix the asymptotic boundary conditions.

{Before proceeding we discuss the notion of ideal contacts. The latter refers
to terminals connected to the edge modes, which absorb the entire impinging current
with no detectable signal away from the contact \cite{Kane_Fisher_1995}. Ideal
contacts have been discussed in Refs.
\onlinecite{Artur_etal_2013,Propotov_Gefen_Mirlin_2017,Nosiglia_etal_2018, Spanslatt_etal_2021} in the absence of interactions, and, in the presence of
interactions, in \onlinecite{Spanslatt_etal_2021}, where it was shown that a
microscopic realization of an ideal contact for counterpropagating edge modes
requires backscattering between them. 
}

Our results can be encapsulated in two ways: Firstly, neutral modes can carry
information about the charge current, information that can be used to reconstruct
the charge current at a different location. Secondly, one can use the charge
chiral mode (top mode of Fig. \ref{fig:geo}) as a ``probe'', and apply it to a
``test'' system (bottom of Fig. \ref{fig:geo}). In this functionality,
our device can be used to detect coherently propagating
bosonic \cite{bosonic} neutral modes in the test system. A
dc charge current at the drain is direct evidence for
neutral modes.

More concretely, let us assume there is at least one left-moving neutral mode
in the test system. When electrons are injected at the
source S if both C$_1$ and C$_2$ are coupled to the modes in region II and IV respectively,
a dc current will be detected at D, regardless of whether the test system has
(right-moving) charge chirals or not.
The presence of right moving chiral charge modes in the test system will not change
this conclusion qualitatively.

Let us understand the physics in
two extreme limits, when (i) both the contacts are coupled, or
(ii) both of the contacts are decoupled.  

{\it Case (i) Both contacts coupled:}  Assuming no charge chiral modes in the test
system, consider a charge (positive by fiat) injected into the probe chiral at
S, which travels to the boundary B$_3$. There, a lump of positive
neutral density (a neutralon) is reflected into the left-moving
neutral mode in region III and lumps of nonuniversal charge are
transmitted into the two right-moving modes in region IV to be fully
absorbed at C$_2$. The left-moving neutralon in III travels to B$_2$,
at which point a positive (electrically) charged lump is reflected into the probe chiral, and an equal and opposite charge is transmitted into the
left-moving mode in the region II, to be fully absorbed at
C$_1$. There will also be a neutralon reflected into the right-moving
neutral chiral in region III, which travels to B$_3$. As usual, this
will undergo transmission and reflection, with the transmitted part
being completely absorbed at C$_2$.  The reflected neutralon part has the same
sign as the original neutralon, and repeats the process described
earlier with a smaller amplitude. With both contacts coupled, an
infinite sequence of charge lumps {\it of the same sign} is detected
at D. Thus, a dc current at S implies a dc current of the
same sign (but with a nonuniversal magnitude) at D.

This is already an instance of the effect we are looking for. Now we add (right-moving) charge chirals to the test system.  All proceeds as
  before until the left-moving neutralon impinges on B$_2$. Now, in addition to the reflected neutral lump, charge lumps will be
  transmitted into the nonuniversal charge modes in region II (to
  be absorbed at C$_1$), and reflected into the probe
  and test charge chirals. The magnitude and sign of the charges are determined by the interaction parameters in region II. Recall that the reflection/transmission is deterministic because no tunneling between the different modes is involved. Thus, there is a dc current at D. 
  
  To summarize, when both contacts are coupled, if a left-moving neutral is present in the test system, there is always a dc current at D, as long as the charge chirals (if any) of the test system are all right-moving.

{\it Case (ii) Both contacts decoupled:} Initially, let us assume that no
charge chiral  modes are present in the test system. The first
step (the injected lump of electric charge traveling from S to B$_3$,
resulting in the reflection of a neutralon and transmission of lumps of
nonuniveral charge  in the two right-moving chirals in region IV) is the
same as before. However, now the right-moving lumps in region IV travel
to B$_4$ and undergo repeated partial reflection and transmission. Similarly,
the left-moving neutralon, upon arriving at B$_2$, results in a charge
lump in the probe charge chiral in III, and a left-moving
charge lump in region II. This latter lump will undergo partial
transmission/reflection at B$_1$. This leads to  multiple scattering
at all the boundaries. However, we can assert, based on charge conservation,
that {\it no dc current is observed at D}. 
Since no left-moving charge modes enter region III, the entire charge
injected at $S$ has to proceed to region V (after multiple scattering
in region IV)
\cite{Safi_Schultz_1995,*Maslov_Stone_1995,*Ponomarenko_1995,*Oreg_Finkelstein_1996}. Any
charge detected at D is initiated by a neutralon arriving at B$_2$ via
the left-moving neutral in III and its descendants via multiple
scattering. Since no total (time-integrated) charge enters  region III from either
of regions II or IV, the time-integrated charge entering the drain D must vanish. Evidently, charge
noise will be detected at D. Similar logic ensures that the dc
charge current exiting region IV into region V is the entire charge
current injected at S.

These conclusions
do not change when we allow (right-moving) charge modes in the test system. Since
the interactions in regions II and region IV are density-density interactions,  the total $U(1)$ 
``charge" (which is completely independent of electric charge) of  each mode has to be conserved in the dc limit.
Thus, we  conclude, that in the presence
of (right-moving) charge chiral modes in the test system, we still need both the
contacts to be coupled in order to have a non-zero current
at the drain D.

In what follows, we will present an outline of the calculations
leading to  our results, relegating straightforward mathematical details
to the supplemental material (SM \cite{SM}). For simplicity, we will
focus on the case where the test system has neutral modes only.

%{\it Reflection coefficients at different boundaries:}
We model the neutrals by an XXZ spin chain and the interaction between
the spin chain and the spin-polarized charged mode as a spin-spin
interaction.  The model is described by the action in
Eq. \ref{eq:action} where the probe charged mode is represented by the
bosonic field $\phi_1$, the right-moving test neutral by $\phi_2$ and
the left-moving test neutral by $\phi_3$. The interaction between the neutrals
$\phi_2$ and $\phi_3$ is denoted by $\lambda_{23}\left( x \right)$.
The interaction between the charged mode and the spin chain, (the same
for both the left- and right-moving neutrals), is denoted by
$\lambda_{12}\left( x \right)$ (= $\lambda_{13}\left( x \right)$)
\begin{align}
S=&\frac{1}{4\pi} \int dxdt \Big[-\partial_x \phi_1 \left(\partial_t \phi_1 
+v_1\partial_x \phi_1\right)\nonumber \\
&-\partial_x \phi_2 \left(\partial_t \phi_2+v_2\partial_x \phi_2\right)+ 
\partial_x \phi_3 \left(\partial_t \phi_3-v_2\partial_x \phi_3\right)\nonumber \\
&-2 \lambda_{12}(x) \partial_x \phi_1 (\partial_x\phi_2+\partial_x\phi_3)-
2 \lambda_{23}(x) \partial_x\phi_2\partial_x\phi_3\Big].
\label{eq:action}
\end{align}

Assuming the interactions are turned on abruptly in regions II and IV,
we calculate the reflection ($r^{\text{B}_\alpha}_{i j}$) and transmission
coefficients ($t^{\text{B}_\alpha}_{i j}$) at B$_2$ and B$_3$, which allows us
to compute the current at D as a function of time via multiple reflections
\cite{Safi_Schultz_1995,*Maslov_Stone_1995,*Ponomarenko_1995,*Oreg_Finkelstein_1996}.

\begin{figure}[h]
\center
\includegraphics[width=\columnwidth]{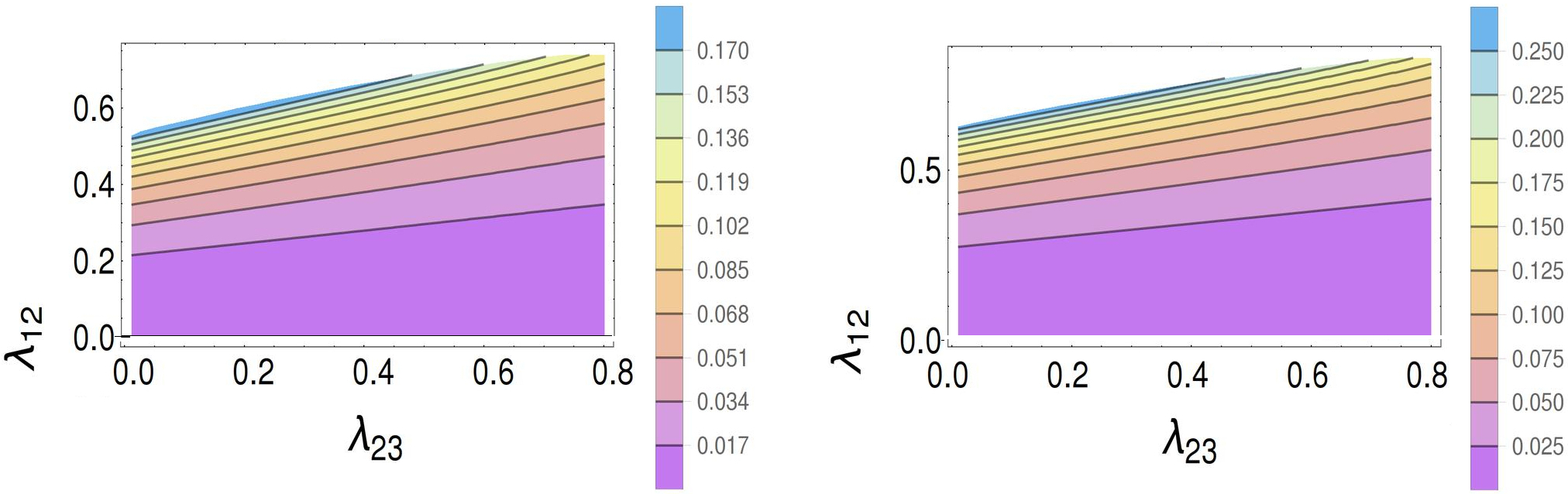}
\caption{The dc current at D as a function of the
$\lambda_{12}$ and $\lambda_{23}$ for two
different values of $v_1, v_2$, when C$_1$ C$_2$ is coupled.
}
\label{fig:current}
\end{figure}

\section{Computation of current and noise}
When both C$_1$ and C$_2$ are coupled, the fraction of the current that
reaches D as a function of time is 
\begin{align}
r_\text{D}(t)=r^{\text{B}_3}_{13}r^{\text{B}_2}_{31}\sum_{n=0}^\infty \left( r^{\text{B}_2}_{32}r^{\text{B}_3}_{23} \right)^n \delta(t-t_{d_n})
\end{align}
where $t_{d_n}=t_0+n \Delta t$. Here $t_0$ is the time for the first
signal and $\Delta t$ is the time for one full reflection
between B$_3$ and B$_2$. The dc current at D is the zero-frequency
limit of the Fourier transform.
\begin{equation}
I_\text{D}(\omega\rightarrow 0)=\frac{r_{13}^{B_3}r_{31}^{B_2}}{1-r_{32}^{B_2}r_{23}^{B_3}}\langle I_\text{tun} \rangle
\label{draincurrent}
\end{equation}
Similarly, we  calculate the noise at D
\cite{Martin_2005,Chamon_Freed_We_1995,Berg_etal_2009} via the
current-current correlation function on a Schwinger-Keldysh contour to obtain
\begin{equation}
N_\text{D}\left(\omega\rightarrow 0\right)=\frac{e(r_{13}^{B_3}r_{31}^{B_2})^2}{1-(r_{32}^{B_3}r_{23}^{B_3})^2}  \langle I_\text{tun} \rangle
\end{equation}
The noise when one or both of the
contacts are decoupled can be computed very similarly \cite{SM}.

%{\it Results:}
When C$_2$ is decoupled the interactions in region IV are purely density-density,
implying  that the $U(1)$ ``charge" of each mode (as previously mentioned, completely
independent of electric charge) is conserved. Thus if we sum up all the multiple
reflections from boundary B$_3$ and B$_4$ (dc limit), the total $U(1)$ ``charge" of the 
neutral reflected from region IV to region III must vanish. Hence the total dc current
at D will be zero. The dc current at the drain is only non-zero if and only if both
the contacts C$_1$ and C$_2$ are coupled.

\section{Experimental Realization}
We now discuss an experimental realization of our setup.  For 
monolayer graphene, Hartree-Fock calculations suggest
\cite{Kharitonov_2012} that at charge neutrality ($\nu=0$), there is
a quantum phase transition between a canted antiferromagnetic (CAF)
phase, stabilized for purely perpendicular magnetic field, and a
spin-polarized phase which can be stabilized by increasing the Zeeman
energy $E_Z$ with an in-plane $B$ field.  The spin-polarized phase has
a fully gapped bulk and a pair of gapless helical edge modes
\cite{Abanin2006,BreyFertig2006}, whereas the CAF phase breaks U(1)
spin-rotation symmetry and has a neutral Goldstone mode in the bulk,
but no gapless charged edge modes
\cite{Ganpathy_Shimshoni_Fertig_2014,MurthyShimshoniFertig2016}.
Experimentally, the phase transition has been seen \cite{Young2014},
but evidence that the phase at purely perpendicular $B$ is the CAF
phase is indirect, via the detection of magnon transmission above
the Zeeman energy \cite{Wei2018}. Indeed, recent scanning tunneling spectroscopy
measurements indicate that the ground state has bond-order
\cite{li2019:stm,liu2021visualizing,coissard2021imaging}. To confirm that the
system has CAF order one would need to detect {\it gapless}
collective excitations, as has been done recently in bilayer graphene
\cite{Hailong_etal_2021}.

A potential experimental realization of the central idea of this
paper is shown in Fig. \ref{fig:prop}. A sheet of graphene in a
perpendicular $B$ field is gated such that the left half is at
filling $\nu=1$, while the right half is at $\nu=0$. In the central
part of the $\nu=0$ region, we overlay graphene with a ferromagnetic
insulator, whose exchange field makes the graphene under it fully polarized and gapped.
However, the annular
periphery of $\nu=0$ region is in the putative CAF phase, with a gapless
Goldstone mode. No topological edge modes exist
between the two phases at $\nu=0$. Confinement in the ``radial''
direction in the $\nu=0$ region will reconstruct the continuum of
bulk Goldstone modes into bands of clockwise-moving and
anticlockwise-moving neutral modes.  The lowest two bands will be
gapless, and represent the counterpropagating neutral modes in
Fig. \ref{fig:geo}.  These  counter-propagating neutral
modes interact with the charge edge mode of the $\nu=1$
quantum Hall phase on the left in the regions where they are proximate (Fig.
\ref{fig:prop}). Adding the source S, drain D, and grounded contact G at
appropriate locations realizes the setup of Fig. \ref{fig:geo}, and provides
a way to unambiguously detect the gapless neutral Goldstone mode of the CAF.  

\begin{figure}[h]
\centering
\includegraphics[width=0.9\columnwidth,angle=0]{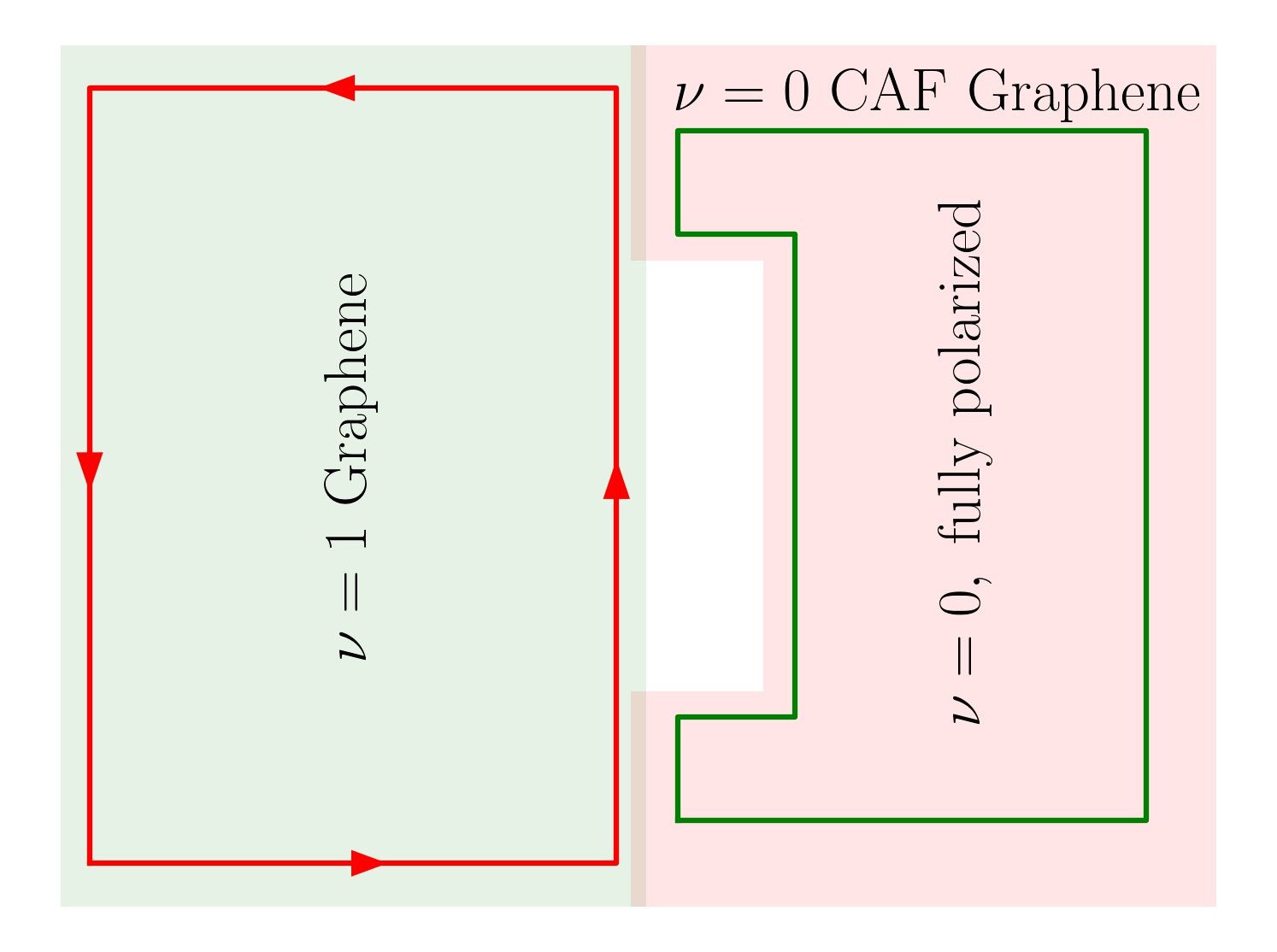}
\caption{A sheet of graphene in a perpendicular $B$-field is gated to have
$\nu=1$ on the left and $\nu=0$ on the right. The central region of $\nu=0$
is overlain by an insulating ferromagnet, inducing the fully polarized phase
of $\nu=0$ graphene in this region. The periphery of $\nu=0$ is presumed to
be in the CAF state, with gapless Goldstone modes. The lowest subband of the
radially confined Goldstone modes interacts with the $\nu=1$ edge mode and
is detected by the scheme described in the text.}
\label{fig:prop}
\end{figure}

\section{Summary and outlook} 
In this work we have
proposed a setup that has two functionalities: (i) Given a system known to have a
neutral mode (the bottom system in Fig. \ref{fig:geo}), we encode information about
the charge current into the neutral current, and subsequently read it
out as a dc charge current at a different spatial location. (ii) Given
a test system suspected of having coherently propagating, bosonic
\cite{bosonic}, neutral modes, we place it along the bottom part of Fig. \ref{fig:geo}
and use our device as a neutral mode detector. An important condition for our
protocol to work is that the neutral mode to be detected should counterpropagate with respect
to all charge modes, else the charge modes will ``short-circuit" the neutral mode.
However, measurements of the upstream and downstream charge conductance along the edge
of the test system are sufficient to determine whether all charge modes co-propagate
in a given system. Backscattering under the contacts breaks the $U(1)$ symmetry
of each mode; without backscattering no dc current at the drain is possible.

Let us elaborate a bit on the functionality of our setup as a neutral mode detector.
Our setup can detect coherently propagating, bosonic \cite{bosonic}, neutral edge
modes when all the charge modes in the test system are gapped, and gapless modes
represent spin/valley fluctuations. Trivial insulators with spontaneous symmetry
breaking of a continuous symmetry, such as the putative CAF phase of graphene at
charge neutrality, are prime examples of such systems. Moreover, our setup will
detect coherently propagating, bosonic, neutral edge modes in  QH systems as well,
as long as two conditions are met: (i) all chiral charge modes of the test QH
system are co-propagating, and (ii) there is at least one neutral mode which
counter-propagates with respect to the charge modes.
For example, the neutral
mode of $\nu=2/3$ at the Kane-Fisher-Polchinski fixed point \cite{KFP_1994} could
be detected by our setup.  Using monolayer graphene for the probe system allows
one to reverse the propagation direction of the probe charge chiral {\it in situ}
by gating to obtain $\nu=\pm1$ in order to realize the geometry of Fig.
\ref{fig:geo}. It must be noted that pairs of neutral edge modes can be generated
by edge reconstructions in quantum Hall systems \cite{SpinModeSwitching,nu4to3}.
We emphasize that our setup can detect coherently propagating bosonic neutral
modes regardless of their physical origin.

Let us compare our setup with previous approaches to neutral mode detection.
In one approach, the passage of upstream  neutral modes through a quantum point
contact was detected through the generation of charge noise
\cite{Bid2010,Bid_etal_2009, Cohen_etal_2019,Biswas_etal_2021}.
More recently, measurements of
heat transport “upstream" as compared to charge transport have been employed
\cite{yacoby_etal_1994,Banerjee_etal_2017,Srivastav_etal_2020}. Not only are
these hard measurements, (they require a precise determination of the
temperature at a given contact), but they cannot determine whether the heat
propagating upstream reflects coherent neutral modes rather than incoherent
transport (e.g., due to diffusive modes). The latter is the result of charge
and heat equilibration \cite{Propotov_Gefen_Mirlin_2017,Nosiglia_etal_2018},
and also leads to upstream charge noise \cite{Spanslatt_etal_2019}.

A second theoretical approach for detecting neutral modes in certain
quantum Hall systems \cite{Feldman_Li_2008,Cano_Nayak_2014} via dc
currents depends on tunneling between QH edges at quantum point contacts, and only
specific neutral modes in specific configurations lead to dc currents.
In our proposal, tunneling between different chiral modes occurs only under the
contacts. 

There are a few unresolved issues of broad import: 
(i) How does one understand
the formulation of linear and non-linear response
in the charge-neutral-charge circuit? 
(ii) Certain exotic spin systems are believed to
have neutral Majorana modes \cite{Kitaev_2006_2,Kasahara_etal_2018}, as is the
$\nu=5/2$ state \cite{Banerjee_etal_2017}. Our proposed
device can detect bosonic \cite{bosonic} neutral modes, but can some extension
thereof be used to detect Majorana modes as well? 

\begin{acknowledgements}
% \section{Acknowledgements :}
We thank A. Mirlin, I. Gornyi, D. Polyakov, and K. Snizhko for their extremely
valuable comments and proposed  modifications  which significantly improved our
manuscript. AD was supported by the German-Israeli Foundation Grant No.
I-1505-303.10/2019 and the GIF. AD also thanks
Israel planning and budgeting committee (PBC) and Weizmann
Institute of Science, Israel Dean of Faculty fellowship,
and Koshland Foundation for financial support.
YG was supported by CRC 183 of the DFG, the Minerva Foundation,
DFG Grant No. MI 658/10-1 and the GIF. SR and GM would like to
thank the VAJRA scheme of SERB, India for its support. GM would
like to thank the US-Israel Binational Science Foundation for its
support via grant no. 2016130, and the Aspen Center for Physics (NSF grant PHY-1607611) where this work was completed.
\end{acknowledgements}
\bibliographystyle{apsrev}
\bibliography{references}

\begin{thebibliography}{44}
\expandafter\ifx\csname natexlab\endcsname\relax\def\natexlab#1{#1}\fi
\expandafter\ifx\csname bibnamefont\endcsname\relax
  \def\bibnamefont#1{#1}\fi
\expandafter\ifx\csname bibfnamefont\endcsname\relax
  \def\bibfnamefont#1{#1}\fi
\expandafter\ifx\csname citenamefont\endcsname\relax
  \def\citenamefont#1{#1}\fi
\expandafter\ifx\csname url\endcsname\relax
  \def\url#1{\texttt{#1}}\fi
\expandafter\ifx\csname urlprefix\endcsname\relax\def\urlprefix{URL }\fi
\providecommand{\bibinfo}[2]{#2}
\providecommand{\eprint}[2][]{\url{#2}}

\bibitem[{\citenamefont{Klitzing et~al.}(1980)\citenamefont{Klitzing, Dorda,
  and Pepper}}]{Klitzing1980}
\bibinfo{author}{\bibfnamefont{K.~v.} \bibnamefont{Klitzing}},
  \bibinfo{author}{\bibfnamefont{G.}~\bibnamefont{Dorda}}, \bibnamefont{and}
  \bibinfo{author}{\bibfnamefont{M.}~\bibnamefont{Pepper}},
  \bibinfo{journal}{Phys. Rev. Lett.} \textbf{\bibinfo{volume}{45}},
  \bibinfo{pages}{494} (\bibinfo{year}{1980}),
  \urlprefix\url{https://link.aps.org/doi/10.1103/PhysRevLett.45.494}.

\bibitem[{\citenamefont{Hasan and Kane}(2010)}]{Hasan2010}
\bibinfo{author}{\bibfnamefont{M.~Z.} \bibnamefont{Hasan}} \bibnamefont{and}
  \bibinfo{author}{\bibfnamefont{C.~L.} \bibnamefont{Kane}},
  \bibinfo{journal}{Rev. Mod. Phys.} \textbf{\bibinfo{volume}{82}},
  \bibinfo{pages}{3045} (\bibinfo{year}{2010}),
  \urlprefix\url{https://link.aps.org/doi/10.1103/RevModPhys.82.3045}.

\bibitem[{\citenamefont{Bid et~al.}(2010)\citenamefont{Bid, Ofek, Inoue,
  Heiblum, Kane, Umansky, and Mahalu}}]{Bid2010}
\bibinfo{author}{\bibfnamefont{A.}~\bibnamefont{Bid}},
  \bibinfo{author}{\bibfnamefont{N.}~\bibnamefont{Ofek}},
  \bibinfo{author}{\bibfnamefont{H.}~\bibnamefont{Inoue}},
  \bibinfo{author}{\bibfnamefont{M.}~\bibnamefont{Heiblum}},
  \bibinfo{author}{\bibfnamefont{C.~L.} \bibnamefont{Kane}},
  \bibinfo{author}{\bibfnamefont{V.}~\bibnamefont{Umansky}}, \bibnamefont{and}
  \bibinfo{author}{\bibfnamefont{D.}~\bibnamefont{Mahalu}},
  \bibinfo{journal}{Nature} \textbf{\bibinfo{volume}{466}},
  \bibinfo{pages}{585} (\bibinfo{year}{2010}), ISSN \bibinfo{issn}{1476-4687},
  \urlprefix\url{https://doi.org/10.1038/nature09277}.

\bibitem[{\citenamefont{Venkatachalam et~al.}(2012)\citenamefont{Venkatachalam,
  Hart, Pfeiffer, West, and Yacoby}}]{Venkatachalam2012}
\bibinfo{author}{\bibfnamefont{V.}~\bibnamefont{Venkatachalam}},
  \bibinfo{author}{\bibfnamefont{S.}~\bibnamefont{Hart}},
  \bibinfo{author}{\bibfnamefont{L.}~\bibnamefont{Pfeiffer}},
  \bibinfo{author}{\bibfnamefont{K.}~\bibnamefont{West}}, \bibnamefont{and}
  \bibinfo{author}{\bibfnamefont{A.}~\bibnamefont{Yacoby}},
  \bibinfo{journal}{Nature Physics} \textbf{\bibinfo{volume}{8}},
  \bibinfo{pages}{676} (\bibinfo{year}{2012}), ISSN \bibinfo{issn}{1745-2481},
  \urlprefix\url{https://doi.org/10.1038/nphys2384}.

\bibitem[{\citenamefont{Deviatov et~al.}(2011)\citenamefont{Deviatov, Lorke,
  Biasiol, and Sorba}}]{Deviatov_etal_2011}
\bibinfo{author}{\bibfnamefont{E.~V.} \bibnamefont{Deviatov}},
  \bibinfo{author}{\bibfnamefont{A.}~\bibnamefont{Lorke}},
  \bibinfo{author}{\bibfnamefont{G.}~\bibnamefont{Biasiol}}, \bibnamefont{and}
  \bibinfo{author}{\bibfnamefont{L.}~\bibnamefont{Sorba}},
  \bibinfo{journal}{Phys. Rev. Lett.} \textbf{\bibinfo{volume}{106}},
  \bibinfo{pages}{256802} (\bibinfo{year}{2011}),
  \urlprefix\url{https://link.aps.org/doi/10.1103/PhysRevLett.106.256802}.

\bibitem[{bos()}]{bosonic}
\bibinfo{note}{By a bosonic mode we mean a chiral mode that can be described by
  a chiral boson living on the edge.}

\bibitem[{\citenamefont{Sp\aa{}nsl\"att
  et~al.}(2021)\citenamefont{Sp\aa{}nsl\"att, Gefen, Gornyi, and
  Polyakov}}]{Spanslatt_etal_2021}
\bibinfo{author}{\bibfnamefont{C.}~\bibnamefont{Sp\aa{}nsl\"att}},
  \bibinfo{author}{\bibfnamefont{Y.}~\bibnamefont{Gefen}},
  \bibinfo{author}{\bibfnamefont{I.~V.} \bibnamefont{Gornyi}},
  \bibnamefont{and} \bibinfo{author}{\bibfnamefont{D.~G.}
  \bibnamefont{Polyakov}}, \bibinfo{journal}{Phys. Rev. B}
  \textbf{\bibinfo{volume}{104}}, \bibinfo{pages}{115416}
  (\bibinfo{year}{2021}),
  \urlprefix\url{https://link.aps.org/doi/10.1103/PhysRevB.104.115416}.

\bibitem[{\citenamefont{Kane and Fisher}(1995)}]{Kane_Fisher_1995}
\bibinfo{author}{\bibfnamefont{C.~L.} \bibnamefont{Kane}} \bibnamefont{and}
  \bibinfo{author}{\bibfnamefont{M.~P.~A.} \bibnamefont{Fisher}},
  \bibinfo{journal}{Phys. Rev. B} \textbf{\bibinfo{volume}{52}},
  \bibinfo{pages}{17393} (\bibinfo{year}{1995}),
  \urlprefix\url{https://link.aps.org/doi/10.1103/PhysRevB.52.17393}.

\bibitem[{\citenamefont{Slobodeniuk et~al.}(2013)\citenamefont{Slobodeniuk,
  Levkivskyi, and Sukhorukov}}]{Artur_etal_2013}
\bibinfo{author}{\bibfnamefont{A.~O.} \bibnamefont{Slobodeniuk}},
  \bibinfo{author}{\bibfnamefont{I.~P.} \bibnamefont{Levkivskyi}},
  \bibnamefont{and} \bibinfo{author}{\bibfnamefont{E.~V.}
  \bibnamefont{Sukhorukov}}, \bibinfo{journal}{Phys. Rev. B}
  \textbf{\bibinfo{volume}{88}}, \bibinfo{pages}{165307}
  (\bibinfo{year}{2013}),
  \urlprefix\url{https://link.aps.org/doi/10.1103/PhysRevB.88.165307}.

\bibitem[{\citenamefont{Protopopov et~al.}(2017)\citenamefont{Protopopov,
  Gefen, and Mirlin}}]{Propotov_Gefen_Mirlin_2017}
\bibinfo{author}{\bibfnamefont{I.}~\bibnamefont{Protopopov}},
  \bibinfo{author}{\bibfnamefont{Y.}~\bibnamefont{Gefen}}, \bibnamefont{and}
  \bibinfo{author}{\bibfnamefont{A.}~\bibnamefont{Mirlin}},
  \bibinfo{journal}{Annals of Physics} \textbf{\bibinfo{volume}{385}},
  \bibinfo{pages}{287} (\bibinfo{year}{2017}), ISSN \bibinfo{issn}{0003-4916},
  \urlprefix\url{https://www.sciencedirect.com/science/article/pii/S0003491617302142}.

\bibitem[{\citenamefont{Nosiglia et~al.}(2018)\citenamefont{Nosiglia, Park,
  Rosenow, and Gefen}}]{Nosiglia_etal_2018}
\bibinfo{author}{\bibfnamefont{C.}~\bibnamefont{Nosiglia}},
  \bibinfo{author}{\bibfnamefont{J.}~\bibnamefont{Park}},
  \bibinfo{author}{\bibfnamefont{B.}~\bibnamefont{Rosenow}}, \bibnamefont{and}
  \bibinfo{author}{\bibfnamefont{Y.}~\bibnamefont{Gefen}},
  \bibinfo{journal}{Phys. Rev. B} \textbf{\bibinfo{volume}{98}},
  \bibinfo{pages}{115408} (\bibinfo{year}{2018}),
  \urlprefix\url{https://link.aps.org/doi/10.1103/PhysRevB.98.115408}.

\bibitem[{\citenamefont{Safi and Schulz}(1995)}]{Safi_Schultz_1995}
\bibinfo{author}{\bibfnamefont{I.}~\bibnamefont{Safi}} \bibnamefont{and}
  \bibinfo{author}{\bibfnamefont{H.~J.} \bibnamefont{Schulz}},
  \bibinfo{journal}{Phys. Rev. B} \textbf{\bibinfo{volume}{52}},
  \bibinfo{pages}{R17040} (\bibinfo{year}{1995}),
  \urlprefix\url{https://link.aps.org/doi/10.1103/PhysRevB.52.R17040}.

\bibitem[{\citenamefont{Maslov and Stone}(1995)}]{Maslov_Stone_1995}
\bibinfo{author}{\bibfnamefont{D.~L.} \bibnamefont{Maslov}} \bibnamefont{and}
  \bibinfo{author}{\bibfnamefont{M.}~\bibnamefont{Stone}},
  \bibinfo{journal}{Phys. Rev. B} \textbf{\bibinfo{volume}{52}},
  \bibinfo{pages}{R5539} (\bibinfo{year}{1995}),
  \urlprefix\url{https://link.aps.org/doi/10.1103/PhysRevB.52.R5539}.

\bibitem[{\citenamefont{Ponomarenko}(1995)}]{Ponomarenko_1995}
\bibinfo{author}{\bibfnamefont{V.~V.} \bibnamefont{Ponomarenko}},
  \bibinfo{journal}{Phys. Rev. B} \textbf{\bibinfo{volume}{52}},
  \bibinfo{pages}{R8666} (\bibinfo{year}{1995}),
  \urlprefix\url{https://link.aps.org/doi/10.1103/PhysRevB.52.R8666}.

\bibitem[{\citenamefont{Oreg and Finkel'stein}(1996)}]{Oreg_Finkelstein_1996}
\bibinfo{author}{\bibfnamefont{Y.}~\bibnamefont{Oreg}} \bibnamefont{and}
  \bibinfo{author}{\bibfnamefont{A.~M.} \bibnamefont{Finkel'stein}},
  \bibinfo{journal}{Phys. Rev. B} \textbf{\bibinfo{volume}{54}},
  \bibinfo{pages}{R14265} (\bibinfo{year}{1996}),
  \urlprefix\url{https://link.aps.org/doi/10.1103/PhysRevB.54.R14265}.

\bibitem[{SM()}]{SM}
\bibinfo{note}{Supplemental Material}.

\bibitem[{\citenamefont{Martin}(2005)}]{Martin_2005}
\bibinfo{author}{\bibfnamefont{T.}~\bibnamefont{Martin}},
  \bibinfo{journal}{Proceedings of the Les Houches Summer School, Session
  LXXXI}  (\bibinfo{year}{2005}).

\bibitem[{\citenamefont{Chamon et~al.}(1995)\citenamefont{Chamon, Freed, and
  Wen}}]{Chamon_Freed_We_1995}
\bibinfo{author}{\bibfnamefont{C.~d.~C.} \bibnamefont{Chamon}},
  \bibinfo{author}{\bibfnamefont{D.~E.} \bibnamefont{Freed}}, \bibnamefont{and}
  \bibinfo{author}{\bibfnamefont{X.~G.} \bibnamefont{Wen}},
  \bibinfo{journal}{Phys. Rev. B} \textbf{\bibinfo{volume}{51}},
  \bibinfo{pages}{2363} (\bibinfo{year}{1995}),
  \urlprefix\url{https://link.aps.org/doi/10.1103/PhysRevB.51.2363}.

\bibitem[{\citenamefont{Berg et~al.}(2009)\citenamefont{Berg, Oreg, Kim, and
  von Oppen}}]{Berg_etal_2009}
\bibinfo{author}{\bibfnamefont{E.}~\bibnamefont{Berg}},
  \bibinfo{author}{\bibfnamefont{Y.}~\bibnamefont{Oreg}},
  \bibinfo{author}{\bibfnamefont{E.-A.} \bibnamefont{Kim}}, \bibnamefont{and}
  \bibinfo{author}{\bibfnamefont{F.}~\bibnamefont{von Oppen}},
  \bibinfo{journal}{Phys. Rev. Lett.} \textbf{\bibinfo{volume}{102}},
  \bibinfo{pages}{236402} (\bibinfo{year}{2009}),
  \urlprefix\url{https://link.aps.org/doi/10.1103/PhysRevLett.102.236402}.

\bibitem[{\citenamefont{Kharitonov}(2012)}]{Kharitonov_2012}
\bibinfo{author}{\bibfnamefont{M.}~\bibnamefont{Kharitonov}},
  \bibinfo{journal}{Phys. Rev. B} \textbf{\bibinfo{volume}{85}},
  \bibinfo{pages}{155439} (\bibinfo{year}{2012}),
  \urlprefix\url{https://link.aps.org/doi/10.1103/PhysRevB.85.155439}.

\bibitem[{\citenamefont{Abanin et~al.}(2006)\citenamefont{Abanin, Lee, and
  Levitov}}]{Abanin2006}
\bibinfo{author}{\bibfnamefont{D.~A.} \bibnamefont{Abanin}},
  \bibinfo{author}{\bibfnamefont{P.~A.} \bibnamefont{Lee}}, \bibnamefont{and}
  \bibinfo{author}{\bibfnamefont{L.~S.} \bibnamefont{Levitov}},
  \bibinfo{journal}{Phys. Rev. Lett.} \textbf{\bibinfo{volume}{96}},
  \bibinfo{pages}{176803} (\bibinfo{year}{2006}),
  \urlprefix\url{https://link.aps.org/doi/10.1103/PhysRevLett.96.176803}.

\bibitem[{\citenamefont{Brey and Fertig}(2006)}]{BreyFertig2006}
\bibinfo{author}{\bibfnamefont{L.}~\bibnamefont{Brey}} \bibnamefont{and}
  \bibinfo{author}{\bibfnamefont{H.~A.} \bibnamefont{Fertig}},
  \bibinfo{journal}{Phys. Rev. B} \textbf{\bibinfo{volume}{73}},
  \bibinfo{pages}{195408} (\bibinfo{year}{2006}),
  \urlprefix\url{https://link.aps.org/doi/10.1103/PhysRevB.73.195408}.

\bibitem[{\citenamefont{Murthy et~al.}(2014)\citenamefont{Murthy, Shimshoni,
  and Fertig}}]{Ganpathy_Shimshoni_Fertig_2014}
\bibinfo{author}{\bibfnamefont{G.}~\bibnamefont{Murthy}},
  \bibinfo{author}{\bibfnamefont{E.}~\bibnamefont{Shimshoni}},
  \bibnamefont{and} \bibinfo{author}{\bibfnamefont{H.~A.}
  \bibnamefont{Fertig}}, \bibinfo{journal}{Phys. Rev. B}
  \textbf{\bibinfo{volume}{90}}, \bibinfo{pages}{241410}
  (\bibinfo{year}{2014}),
  \urlprefix\url{https://link.aps.org/doi/10.1103/PhysRevB.90.241410}.

\bibitem[{\citenamefont{Murthy et~al.}(2016)\citenamefont{Murthy, Shimshoni,
  and Fertig}}]{MurthyShimshoniFertig2016}
\bibinfo{author}{\bibfnamefont{G.}~\bibnamefont{Murthy}},
  \bibinfo{author}{\bibfnamefont{E.}~\bibnamefont{Shimshoni}},
  \bibnamefont{and} \bibinfo{author}{\bibfnamefont{H.~A.}
  \bibnamefont{Fertig}}, \bibinfo{journal}{Phys. Rev. B}
  \textbf{\bibinfo{volume}{93}}, \bibinfo{pages}{045105}
  (\bibinfo{year}{2016}),
  \urlprefix\url{https://link.aps.org/doi/10.1103/PhysRevB.93.045105}.

\bibitem[{\citenamefont{Young et~al.}(2014)\citenamefont{Young,
  Sanchez-Yamagishi, Hunt, Choi, Watanabe, Taniguchi, Ashoori, and
  Jarillo-Herrero}}]{Young2014}
\bibinfo{author}{\bibfnamefont{A.~F.} \bibnamefont{Young}},
  \bibinfo{author}{\bibfnamefont{J.~D.} \bibnamefont{Sanchez-Yamagishi}},
  \bibinfo{author}{\bibfnamefont{B.}~\bibnamefont{Hunt}},
  \bibinfo{author}{\bibfnamefont{S.~H.} \bibnamefont{Choi}},
  \bibinfo{author}{\bibfnamefont{K.}~\bibnamefont{Watanabe}},
  \bibinfo{author}{\bibfnamefont{T.}~\bibnamefont{Taniguchi}},
  \bibinfo{author}{\bibfnamefont{R.~C.} \bibnamefont{Ashoori}},
  \bibnamefont{and}
  \bibinfo{author}{\bibfnamefont{P.}~\bibnamefont{Jarillo-Herrero}},
  \bibinfo{journal}{Nature} \textbf{\bibinfo{volume}{505}},
  \bibinfo{pages}{528} (\bibinfo{year}{2014}), ISSN \bibinfo{issn}{1476-4687},
  \urlprefix\url{https://doi.org/10.1038/nature12800}.

\bibitem[{\citenamefont{Wei et~al.}(2018)\citenamefont{Wei, van~der Sar, Lee,
  Watanabe, Taniguchi, Halperin, and Yacoby}}]{Wei2018}
\bibinfo{author}{\bibfnamefont{D.~S.} \bibnamefont{Wei}},
  \bibinfo{author}{\bibfnamefont{T.}~\bibnamefont{van~der Sar}},
  \bibinfo{author}{\bibfnamefont{S.~H.} \bibnamefont{Lee}},
  \bibinfo{author}{\bibfnamefont{K.}~\bibnamefont{Watanabe}},
  \bibinfo{author}{\bibfnamefont{T.}~\bibnamefont{Taniguchi}},
  \bibinfo{author}{\bibfnamefont{B.~I.} \bibnamefont{Halperin}},
  \bibnamefont{and} \bibinfo{author}{\bibfnamefont{A.}~\bibnamefont{Yacoby}},
  \bibinfo{journal}{Science} \textbf{\bibinfo{volume}{362}},
  \bibinfo{pages}{229} (\bibinfo{year}{2018}),
  \eprint{https://science.sciencemag.org/content/362/6411/229.full.pdf},
  \urlprefix\url{https://science.sciencemag.org/content/362/6411/229}.

\bibitem[{\citenamefont{Li et~al.}(2019)\citenamefont{Li, Zhang, Yin, and
  He}}]{li2019:stm}
\bibinfo{author}{\bibfnamefont{S.-Y.} \bibnamefont{Li}},
  \bibinfo{author}{\bibfnamefont{Y.}~\bibnamefont{Zhang}},
  \bibinfo{author}{\bibfnamefont{L.-J.} \bibnamefont{Yin}}, \bibnamefont{and}
  \bibinfo{author}{\bibfnamefont{L.}~\bibnamefont{He}}, \bibinfo{journal}{Phys.
  Rev. B} \textbf{\bibinfo{volume}{100}}, \bibinfo{pages}{085437}
  (\bibinfo{year}{2019}),
  \urlprefix\url{https://link.aps.org/doi/10.1103/PhysRevB.100.085437}.

\bibitem[{\citenamefont{Liu et~al.}(2022)\citenamefont{Liu, Farahi, Chiu,
  Papic, Watanabe, Taniguchi, Zaletel, and Yazdani}}]{liu2021visualizing}
\bibinfo{author}{\bibfnamefont{X.}~\bibnamefont{Liu}},
  \bibinfo{author}{\bibfnamefont{G.}~\bibnamefont{Farahi}},
  \bibinfo{author}{\bibfnamefont{C.-L.} \bibnamefont{Chiu}},
  \bibinfo{author}{\bibfnamefont{Z.}~\bibnamefont{Papic}},
  \bibinfo{author}{\bibfnamefont{K.}~\bibnamefont{Watanabe}},
  \bibinfo{author}{\bibfnamefont{T.}~\bibnamefont{Taniguchi}},
  \bibinfo{author}{\bibfnamefont{M.~P.} \bibnamefont{Zaletel}},
  \bibnamefont{and} \bibinfo{author}{\bibfnamefont{A.}~\bibnamefont{Yazdani}},
  \bibinfo{journal}{Science} \textbf{\bibinfo{volume}{375}},
  \bibinfo{pages}{321} (\bibinfo{year}{2022}),
  \eprint{https://www.science.org/doi/pdf/10.1126/science.abm3770},
  \urlprefix\url{https://www.science.org/doi/abs/10.1126/science.abm3770}.

\bibitem[{\citenamefont{Coissard et~al.}(2021)\citenamefont{Coissard, Wander,
  Vignaud, Grushin, Repellin, Watanabe, Taniguchi, Gay, Winkelmann, Courtois
  et~al.}}]{coissard2021imaging}
\bibinfo{author}{\bibfnamefont{A.}~\bibnamefont{Coissard}},
  \bibinfo{author}{\bibfnamefont{D.}~\bibnamefont{Wander}},
  \bibinfo{author}{\bibfnamefont{H.}~\bibnamefont{Vignaud}},
  \bibinfo{author}{\bibfnamefont{A.~G.} \bibnamefont{Grushin}},
  \bibinfo{author}{\bibfnamefont{C.}~\bibnamefont{Repellin}},
  \bibinfo{author}{\bibfnamefont{K.}~\bibnamefont{Watanabe}},
  \bibinfo{author}{\bibfnamefont{T.}~\bibnamefont{Taniguchi}},
  \bibinfo{author}{\bibfnamefont{F.}~\bibnamefont{Gay}},
  \bibinfo{author}{\bibfnamefont{C.}~\bibnamefont{Winkelmann}},
  \bibinfo{author}{\bibfnamefont{H.}~\bibnamefont{Courtois}},
  \bibnamefont{et~al.}, \emph{\bibinfo{title}{Imaging tunable quantum hall
  broken-symmetry orders in charge-neutral graphene}} (\bibinfo{year}{2021}),
  \eprint{2110.02811}.

\bibitem[{\citenamefont{Fu et~al.}(2021)\citenamefont{Fu, Huang, Watanabe,
  Taniguchi, and Zhu}}]{Hailong_etal_2021}
\bibinfo{author}{\bibfnamefont{H.}~\bibnamefont{Fu}},
  \bibinfo{author}{\bibfnamefont{K.}~\bibnamefont{Huang}},
  \bibinfo{author}{\bibfnamefont{K.}~\bibnamefont{Watanabe}},
  \bibinfo{author}{\bibfnamefont{T.}~\bibnamefont{Taniguchi}},
  \bibnamefont{and} \bibinfo{author}{\bibfnamefont{J.}~\bibnamefont{Zhu}},
  \bibinfo{journal}{Phys. Rev. X} \textbf{\bibinfo{volume}{11}},
  \bibinfo{pages}{021012} (\bibinfo{year}{2021}),
  \urlprefix\url{https://link.aps.org/doi/10.1103/PhysRevX.11.021012}.

\bibitem[{\citenamefont{Kane et~al.}(1994)\citenamefont{Kane, Fisher, and
  Polchinski}}]{KFP_1994}
\bibinfo{author}{\bibfnamefont{C.~L.} \bibnamefont{Kane}},
  \bibinfo{author}{\bibfnamefont{M.~P.~A.} \bibnamefont{Fisher}},
  \bibnamefont{and}
  \bibinfo{author}{\bibfnamefont{J.}~\bibnamefont{Polchinski}},
  \bibinfo{journal}{Phys. Rev. Lett.} \textbf{\bibinfo{volume}{72}},
  \bibinfo{pages}{4129} (\bibinfo{year}{1994}),
  \urlprefix\url{https://link.aps.org/doi/10.1103/PhysRevLett.72.4129}.

\bibitem[{\citenamefont{Khanna et~al.}(2017)\citenamefont{Khanna, Murthy, Rao,
  and Gefen}}]{SpinModeSwitching}
\bibinfo{author}{\bibfnamefont{U.}~\bibnamefont{Khanna}},
  \bibinfo{author}{\bibfnamefont{G.}~\bibnamefont{Murthy}},
  \bibinfo{author}{\bibfnamefont{S.}~\bibnamefont{Rao}}, \bibnamefont{and}
  \bibinfo{author}{\bibfnamefont{Y.}~\bibnamefont{Gefen}},
  \bibinfo{journal}{Phys. Rev. Lett.} \textbf{\bibinfo{volume}{119}},
  \bibinfo{pages}{186804} (\bibinfo{year}{2017}),
  \urlprefix\url{https://link.aps.org/doi/10.1103/PhysRevLett.119.186804}.

\bibitem[{\citenamefont{Saha et~al.}(2021)\citenamefont{Saha, De, Rao, Gefen,
  and Murthy}}]{nu4to3}
\bibinfo{author}{\bibfnamefont{A.}~\bibnamefont{Saha}},
  \bibinfo{author}{\bibfnamefont{S.~J.} \bibnamefont{De}},
  \bibinfo{author}{\bibfnamefont{S.}~\bibnamefont{Rao}},
  \bibinfo{author}{\bibfnamefont{Y.}~\bibnamefont{Gefen}}, \bibnamefont{and}
  \bibinfo{author}{\bibfnamefont{G.}~\bibnamefont{Murthy}},
  \bibinfo{journal}{Phys. Rev. B} \textbf{\bibinfo{volume}{103}},
  \bibinfo{pages}{L081401} (\bibinfo{year}{2021}),
  \urlprefix\url{https://link.aps.org/doi/10.1103/PhysRevB.103.L081401}.

\bibitem[{\citenamefont{Bid et~al.}(2009)\citenamefont{Bid, Ofek, Heiblum,
  Umansky, and Mahalu}}]{Bid_etal_2009}
\bibinfo{author}{\bibfnamefont{A.}~\bibnamefont{Bid}},
  \bibinfo{author}{\bibfnamefont{N.}~\bibnamefont{Ofek}},
  \bibinfo{author}{\bibfnamefont{M.}~\bibnamefont{Heiblum}},
  \bibinfo{author}{\bibfnamefont{V.}~\bibnamefont{Umansky}}, \bibnamefont{and}
  \bibinfo{author}{\bibfnamefont{D.}~\bibnamefont{Mahalu}},
  \bibinfo{journal}{Phys. Rev. Lett.} \textbf{\bibinfo{volume}{103}},
  \bibinfo{pages}{236802} (\bibinfo{year}{2009}),
  \urlprefix\url{https://link.aps.org/doi/10.1103/PhysRevLett.103.236802}.

\bibitem[{\citenamefont{Cohen et~al.}(2019)\citenamefont{Cohen, Ronen, Yang,
  Banitt, Park, Heiblum, Mirlin, Gefen, and Umansky}}]{Cohen_etal_2019}
\bibinfo{author}{\bibfnamefont{Y.}~\bibnamefont{Cohen}},
  \bibinfo{author}{\bibfnamefont{Y.}~\bibnamefont{Ronen}},
  \bibinfo{author}{\bibfnamefont{W.}~\bibnamefont{Yang}},
  \bibinfo{author}{\bibfnamefont{D.}~\bibnamefont{Banitt}},
  \bibinfo{author}{\bibfnamefont{J.}~\bibnamefont{Park}},
  \bibinfo{author}{\bibfnamefont{M.}~\bibnamefont{Heiblum}},
  \bibinfo{author}{\bibfnamefont{A.~D.} \bibnamefont{Mirlin}},
  \bibinfo{author}{\bibfnamefont{Y.}~\bibnamefont{Gefen}}, \bibnamefont{and}
  \bibinfo{author}{\bibfnamefont{V.}~\bibnamefont{Umansky}},
  \bibinfo{journal}{Nature Communications} \textbf{\bibinfo{volume}{10}},
  \bibinfo{pages}{1920} (\bibinfo{year}{2019}), ISSN \bibinfo{issn}{2041-1723},
  \urlprefix\url{https://doi.org/10.1038/s41467-019-09920-5}.

\bibitem[{\citenamefont{Biswas et~al.}()\citenamefont{Biswas, Das, Kundu,
  Gefen, and Heiblum}}]{Biswas_etal_2021}
\bibinfo{author}{\bibfnamefont{S.}~\bibnamefont{Biswas}},
  \bibinfo{author}{\bibfnamefont{A.}~\bibnamefont{Das}},
  \bibinfo{author}{\bibfnamefont{H.~K.} \bibnamefont{Kundu}},
  \bibinfo{author}{\bibfnamefont{Y.}~\bibnamefont{Gefen}}, \bibnamefont{and}
  \bibinfo{author}{\bibfnamefont{M.}~\bibnamefont{Heiblum}},
  \emph{\bibinfo{title}{in preparation}}.

\bibitem[{\citenamefont{Yacoby et~al.}(1994)\citenamefont{Yacoby, Heiblum,
  Shtrikman, Umansky, and Mahalu}}]{yacoby_etal_1994}
\bibinfo{author}{\bibfnamefont{A.}~\bibnamefont{Yacoby}},
  \bibinfo{author}{\bibfnamefont{M.}~\bibnamefont{Heiblum}},
  \bibinfo{author}{\bibfnamefont{H.}~\bibnamefont{Shtrikman}},
  \bibinfo{author}{\bibfnamefont{V.}~\bibnamefont{Umansky}}, \bibnamefont{and}
  \bibinfo{author}{\bibfnamefont{D.}~\bibnamefont{Mahalu}},
  \bibinfo{journal}{Semiconductor Science and Technology}
  \textbf{\bibinfo{volume}{9}}, \bibinfo{pages}{907} (\bibinfo{year}{1994}).

\bibitem[{\citenamefont{Banerjee et~al.}(2017)\citenamefont{Banerjee, Heiblum,
  Rosenblatt, Oreg, Feldman, Stern, and Umansky}}]{Banerjee_etal_2017}
\bibinfo{author}{\bibfnamefont{M.}~\bibnamefont{Banerjee}},
  \bibinfo{author}{\bibfnamefont{M.}~\bibnamefont{Heiblum}},
  \bibinfo{author}{\bibfnamefont{A.}~\bibnamefont{Rosenblatt}},
  \bibinfo{author}{\bibfnamefont{Y.}~\bibnamefont{Oreg}},
  \bibinfo{author}{\bibfnamefont{D.~E.} \bibnamefont{Feldman}},
  \bibinfo{author}{\bibfnamefont{A.}~\bibnamefont{Stern}}, \bibnamefont{and}
  \bibinfo{author}{\bibfnamefont{V.}~\bibnamefont{Umansky}},
  \bibinfo{journal}{Nature} \textbf{\bibinfo{volume}{545}}, \bibinfo{pages}{75}
  (\bibinfo{year}{2017}), ISSN \bibinfo{issn}{1476-4687},
  \urlprefix\url{https://doi.org/10.1038/nature22052}.

\bibitem[{\citenamefont{Srivastav et~al.}(2021)\citenamefont{Srivastav, Kumar,
  Sp\aa{}nsl\"att, Watanabe, Taniguchi, Mirlin, Gefen, and
  Das}}]{Srivastav_etal_2020}
\bibinfo{author}{\bibfnamefont{S.~K.} \bibnamefont{Srivastav}},
  \bibinfo{author}{\bibfnamefont{R.}~\bibnamefont{Kumar}},
  \bibinfo{author}{\bibfnamefont{C.}~\bibnamefont{Sp\aa{}nsl\"att}},
  \bibinfo{author}{\bibfnamefont{K.}~\bibnamefont{Watanabe}},
  \bibinfo{author}{\bibfnamefont{T.}~\bibnamefont{Taniguchi}},
  \bibinfo{author}{\bibfnamefont{A.~D.} \bibnamefont{Mirlin}},
  \bibinfo{author}{\bibfnamefont{Y.}~\bibnamefont{Gefen}}, \bibnamefont{and}
  \bibinfo{author}{\bibfnamefont{A.}~\bibnamefont{Das}},
  \bibinfo{journal}{Phys. Rev. Lett.} \textbf{\bibinfo{volume}{126}},
  \bibinfo{pages}{216803} (\bibinfo{year}{2021}),
  \urlprefix\url{https://link.aps.org/doi/10.1103/PhysRevLett.126.216803}.

\bibitem[{\citenamefont{Sp\aa{}nsl\"att
  et~al.}(2019)\citenamefont{Sp\aa{}nsl\"att, Park, Gefen, and
  Mirlin}}]{Spanslatt_etal_2019}
\bibinfo{author}{\bibfnamefont{C.}~\bibnamefont{Sp\aa{}nsl\"att}},
  \bibinfo{author}{\bibfnamefont{J.}~\bibnamefont{Park}},
  \bibinfo{author}{\bibfnamefont{Y.}~\bibnamefont{Gefen}}, \bibnamefont{and}
  \bibinfo{author}{\bibfnamefont{A.~D.} \bibnamefont{Mirlin}},
  \bibinfo{journal}{Phys. Rev. Lett.} \textbf{\bibinfo{volume}{123}},
  \bibinfo{pages}{137701} (\bibinfo{year}{2019}),
  \urlprefix\url{https://link.aps.org/doi/10.1103/PhysRevLett.123.137701}.

\bibitem[{\citenamefont{Feldman and Li}(2008)}]{Feldman_Li_2008}
\bibinfo{author}{\bibfnamefont{D.~E.} \bibnamefont{Feldman}} \bibnamefont{and}
  \bibinfo{author}{\bibfnamefont{F.}~\bibnamefont{Li}}, \bibinfo{journal}{Phys.
  Rev. B} \textbf{\bibinfo{volume}{78}}, \bibinfo{pages}{161304}
  (\bibinfo{year}{2008}),
  \urlprefix\url{https://link.aps.org/doi/10.1103/PhysRevB.78.161304}.

\bibitem[{\citenamefont{Cano and Nayak}(2014)}]{Cano_Nayak_2014}
\bibinfo{author}{\bibfnamefont{J.}~\bibnamefont{Cano}} \bibnamefont{and}
  \bibinfo{author}{\bibfnamefont{C.}~\bibnamefont{Nayak}},
  \bibinfo{journal}{Phys. Rev. B} \textbf{\bibinfo{volume}{90}},
  \bibinfo{pages}{235109} (\bibinfo{year}{2014}),
  \urlprefix\url{https://link.aps.org/doi/10.1103/PhysRevB.90.235109}.

\bibitem[{\citenamefont{Kitaev}(2006)}]{Kitaev_2006_2}
\bibinfo{author}{\bibfnamefont{A.}~\bibnamefont{Kitaev}},
  \bibinfo{journal}{Annals of Physics} \textbf{\bibinfo{volume}{321}},
  \bibinfo{pages}{2} (\bibinfo{year}{2006}), ISSN \bibinfo{issn}{0003-4916},
  \bibinfo{note}{january Special Issue},
  \urlprefix\url{https://www.sciencedirect.com/science/article/pii/S0003491605002381}.

\bibitem[{\citenamefont{Kasahara et~al.}(2018)\citenamefont{Kasahara, Ohnishi,
  Mizukami, Tanaka, Ma, Sugii, Kurita, Tanaka, Nasu, Motome
  et~al.}}]{Kasahara_etal_2018}
\bibinfo{author}{\bibfnamefont{Y.}~\bibnamefont{Kasahara}},
  \bibinfo{author}{\bibfnamefont{T.}~\bibnamefont{Ohnishi}},
  \bibinfo{author}{\bibfnamefont{Y.}~\bibnamefont{Mizukami}},
  \bibinfo{author}{\bibfnamefont{O.}~\bibnamefont{Tanaka}},
  \bibinfo{author}{\bibfnamefont{S.}~\bibnamefont{Ma}},
  \bibinfo{author}{\bibfnamefont{K.}~\bibnamefont{Sugii}},
  \bibinfo{author}{\bibfnamefont{N.}~\bibnamefont{Kurita}},
  \bibinfo{author}{\bibfnamefont{H.}~\bibnamefont{Tanaka}},
  \bibinfo{author}{\bibfnamefont{J.}~\bibnamefont{Nasu}},
  \bibinfo{author}{\bibfnamefont{Y.}~\bibnamefont{Motome}},
  \bibnamefont{et~al.}, \bibinfo{journal}{Nature}
  \textbf{\bibinfo{volume}{559}}, \bibinfo{pages}{227} (\bibinfo{year}{2018}),
  ISSN \bibinfo{issn}{1476-4687},
  \urlprefix\url{https://doi.org/10.1038/s41586-018-0274-0}.

\end{thebibliography}

\newpage

\clearpage

\newpage\newpage



\section*{Supplementary material}

# Supplement to Dc Electrical Current Generated by Upstream Neutral Modes

Ankur Das,[1, *] Sumathi Rao,[2] Yuval Gefen,[1] and Ganpathy Murthy[3]

[1] *Department of Condensed Matter Physics, Weizmann Institute of Science, Rehovot, 76100 Israel*
[2] *Harish-Chandra Research Institute, HBNI, Chhatnag Road, Jhunsi, Allahabad 211 019, India*
[3] *Department of Physics and Astronomy, University of Kentucky, Lexington, KY 40506, USA*

Here we describe the details of the calculation in supplement to the main text.

In this set of supplemental materials, we provide details of the action (Section SI), how the reflection and transmission coefficients are computed (Section SII), and how the current and the current noise at the drain are computed (Section SIII). While most of our analysis is in the case when the test system contains neutral modes only, we also analyze (Section SIV) an interesting special case when the "test" system (bottom of Fig. 1 in the main text) is the fractional quantum Hall state at $\nu = 2/3$, and thus has charge as well as neutral edge modes.

## SI. ACTION AND EIGENMODES

In this section, we present details of the case considered in the main text, which assumes that there are no charge modes in the test system.

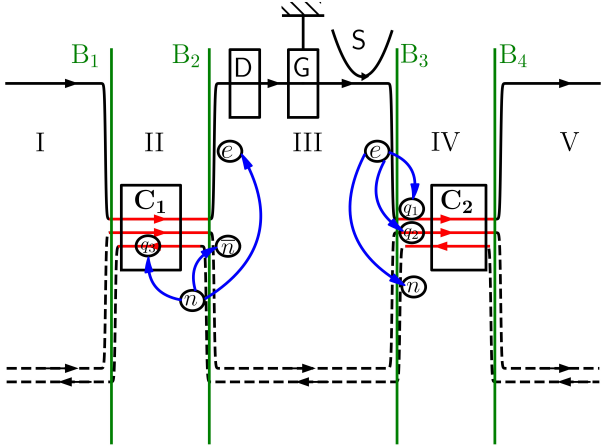

FIG. S1. The geometry of our detector, for the case, when the test system has no charge chiral edge modes. In regions II and IV there are density-density interactions between the test charge chiral and the test neutral chirals. $C_1$ and $C_2$ are perfect ohmic contacts while S, D are the source and drain separated by the grounded contact G.

---

* ankur.das@weizmann.ac.il

The action corresponding to our model in Fig. S1 is

$$S = \frac{1}{4\pi} \int dx dt \Big[ -\partial_x \phi_1 (\partial_t \phi_1 + v_1 \partial_x \phi_1) \\
- \partial_x \phi_2 (\partial_t \phi_2 + v_2 \partial_x \phi_2) + \partial_x \phi_3 (\partial_t \phi_3 - v_2 \partial_x \phi_3) \\
- 2\lambda_{12}(x) \partial_x \phi_1 (\partial_x \phi_2 + \partial_x \phi_3) - 2\lambda_{23}(x) \partial_x \phi_2 \partial_x \phi_3 \Big]. \tag{S1}$$

where we model the neutrals by an XXZ spin chain. Here $\phi_1$ is the charge mode, $\phi_2$ is the right-moving neutral mode and $\phi_3$ is the left-moving neutral mode. The interaction between the neutrals, denoted by $\lambda_{23}(x) = \lambda_{23}$, is a constant everywhere. The interaction parameters between the charged mode and the neutral spin chain modes are $\lambda_{12}(x)$ ($= \lambda_{13}(x)$), and are nonzero only in regions II and IV, and switch on abruptly at the boundaries of regions II and IV.

We define eigenmodes in each region, depending on the interaction strengths, as a linear combination of the bare modes. In regions where the test charged chiral mode is coupled (regions of II and IV) we write the bare fields $\phi_\alpha^i$ in terms of the eigenmodes of Eq. S1 $\tilde{\phi}_\alpha^i$ as,

$$\phi_\alpha^i = M_{\alpha\beta}\, \tilde{\phi}_\beta^i. \tag{S2}$$

Similarly for region $j =$ I, III we can write the bare fields in terms of the eigenmodes $(\phi_\beta^j)$ as,

$$\phi_\alpha^j = N_{\alpha\beta}\, \tilde{\phi}_\beta^j. \tag{S3}$$

We will use these definitions to calculate the reflection and transmission coefficients at every boundary. The matrix $N$ depends on a single parameter because only the two bare neutral modes of the test system are mixed. However, the matrix $M$ is more complex, and can be written as a real member of the group $SO(2,1)$. More explicitly,

$$M = \begin{bmatrix} 1 & 0 & 0 \\ 0 & \cosh(\xi_1) & \sinh(\xi_1) \\ 0 & \sinh(\xi_1) & \cosh(\xi_1) \end{bmatrix} \cdot \begin{bmatrix} \cosh(\xi_2) & 0 & \sinh(\xi_2) \\ 0 & 1 & 0 \\ \sinh(\xi_2) & 0 & \cosh(\xi_2) \end{bmatrix}$$
$$\cdot \begin{bmatrix} \cos(\theta) & \sin(\theta) & 0 \\ -\sin(\theta) & \cos(\theta) & 0 \\ 0 & 0 & 1 \end{bmatrix} \tag{S4a}$$

$$N = \begin{bmatrix} 1 & 0 & 0 \\ 0 & \cosh(\xi) & \sinh(\xi) \\ 0 & \sinh(\xi) & \cosh(\xi) \end{bmatrix}. \tag{S4b}$$

The $\theta, \xi_i$ appearing in these matrices can be determined in a straightforward manner from the action.

## SII. FINDING REFLECTION AND TRANSMISSION

The equations of motion resulting from the action of Eq. S1 are

$$\frac{\partial^2 \phi_1}{\partial x \partial t} + v_1 \frac{\partial^2 \phi_1}{\partial x^2} + \frac{d}{dx}\left[\lambda_{12}(x)\left(\frac{\partial \phi_2}{\partial x} + \frac{\partial \phi_3}{\partial x}\right)\right] = 0 \tag{S5a}$$

$$\frac{\partial^2 \phi_2}{\partial x \partial t} + v_2 \frac{\partial^2 \phi_2}{\partial x^2} + \frac{d}{dx}\left[\lambda_{12}(x)\frac{\partial \phi_1}{\partial x} + \lambda_{23}\frac{\partial \phi_3}{\partial x}\right] = 0 \tag{S5b}$$

$$\frac{\partial^2 \phi_3}{\partial x \partial t} - v_3 \frac{\partial^2 \phi_2}{\partial x^2} - \frac{d}{dx}\left[\lambda_{12}(x)\frac{\partial \phi_1}{\partial x} + \lambda_{23}\frac{\partial \phi_3}{\partial x}\right] = 0. \tag{S5c}$$

For future convenience, we have kept the notations $v_2$ and $v_3$, though for the purposes of this section $v_2 = v_3$. Fourier transforming them to the frequency domain results in the following equations.

$$-i\omega\phi_1' + v_1\phi_1'' + [\lambda_{12}(x)(\phi_2' + \phi_3')]' = 0 \tag{S6a}$$
$$-i\omega\phi_2' + v_2\phi_2'' + [\lambda_{12}(x)\phi_1' + \lambda_{23}\phi_3']' = 0 \tag{S6b}$$
$$-i\omega\phi_3' - v_3\phi_3'' - [\lambda_{12}(x)\phi_1' + \lambda_{23}\phi_2']' = 0. \tag{S6c}$$

Here prime represents derivative with respect to $x$. Now we integrate across the boundary, assuming $\phi_i$ to be continuous across it. This leads to the following boundary conditions across an arbitrary boundary where the interaction strengths change abruptly:

$$v_1 [\phi_1']_{0-}^{0+} + [\lambda_{12}(x)(\phi_2' + \phi_3')]_{0-}^{0+} = 0 \tag{S7a}$$
$$v_2 [\phi_2']_{0-}^{0+} + [\lambda_{12}(x)\phi_1' + \lambda_{23}\phi_3']_{0-}^{0+} = 0 \tag{S7b}$$
$$v_3 [\phi_2']_{0-}^{0+} + [\lambda_{12}(x)\phi_1' + \lambda_{23}\phi_2']_{0-}^{0+} = 0. \tag{S7c}$$

Here $0\pm$ represents across the boundary, and depending on the boundary, the values of $\lambda_{ij}$ change.

### A. Boundary type I

This type of boundary corresponds to the probe chiral not interacting with anything to the left, while it interacts with the test chirals on the right, as in the cases of boundaries $B_1$ and $B_3$. If the boundary is at $x = 0$ we have

$$\lambda_{12}(x) = \lambda_{12}\ \theta(x). \tag{S8}$$

We now write the boundary conditions for the three incoming channels as

Case I: Source coming from the left through channel 1, i.e. $\left(\tilde{\phi}_1^j\right)'\Big|_{0-} = \left(\phi_1^j\right)'\Big|_{0-} = 1$, and

$$\left(\tilde{\phi}_2^j\right)'\Big|_{0-} = N_{2\beta}^{-1}\left(\phi_\beta^j\right)'\Big|_{0-} = 0 \tag{S9a}$$

$$\left(\tilde{\phi}_3^i\right)'\Big|_{0+} = M_{3\beta}^{-1}\left(\phi_\beta^i\right)'\Big|_{0+} = 0 \tag{S9b}$$

Using these boundary conditions with Eq. S7 we solve for the reflection and transmission coefficients. We will denote the transmission coefficient ($t_{\alpha,\beta}^{B_j}$) and reflection coefficients ($r_{\alpha,\beta}^{B_j}$) where the respective transmission and reflection happens from $\alpha$ mode to $\beta$ mode at boundary $B_j$. We will use this notation from now on.

Case II: Source coming from the left through channel 2, i.e. $\left(\tilde{\phi}_2^j\right)'\Big|_{0-} = N_{2\beta}^{-1}\left(\phi_\beta^j\right)'\Big|_{0-} = 1$ and

$$\left(\tilde{\phi}_1^j\right)'\Big|_{0-} = \left(\phi_1^j\right)'\Big|_{0-} = 0 \tag{S10a}$$

$$\left(\tilde{\phi}_3^i\right)'\Big|_{0+} = M_{3\beta}^{-1}\left(\phi_\beta^i\right)'\Big|_{0+} = 0 \tag{S10b}$$

Case III: Source coming from the right through channel 3, i.e. $\left(\tilde{\phi}_3^i\right)'\Big|_{0+} = M_{3\beta}^{-1}\left(\phi_\beta^i\right)'\Big|_{0+} = 1$ and

$$\left(\tilde{\phi}_1^j\right)'\Big|_{0-} = \left(\phi_1^j\right)'\Big|_{0-} = 0 \tag{S11a}$$

$$\left(\tilde{\phi}_2^j\right)'\Big|_{0-} = N_{2\beta}^{-1}\left(\phi_\beta^j\right)'\Big|_{0-} = 0 \tag{S11b}$$

### B. Boundary type II

In this type of boundary, the probe chiral interacts with the test chirals for $x < 0$, but does not interact for $x > 0$. Examples are $B_2$ and $B_4$, where we have

$$\lambda_{12}(x) = \lambda_{12}\ \theta(-x). \tag{S12}$$

For this let us again write down different boundary conditions,

Case I: Source coming from the left through channel 1, i.e. $\left(\tilde{\phi}_1^j\right)'\Big|_{0-} = M_{1\beta}^{-1}\left(\phi_\beta^j\right)'\Big|_{0-} = 1$ and

$$\left(\tilde{\phi}_2^i\right)'\Big|_{0-} = M_{2\beta}^{-1}\left(\phi_\beta^i\right)'\Big|_{0-} = 0 \tag{S13a}$$

$$\left(\tilde{\phi}_3^j\right)'\Big|_{0+} = N_{3\beta}^{-1}\left(\phi_\beta^j\right)'\Big|_{0+} = 0 \tag{S13b}$$

Case II: Source coming from the left through channel 2, i.e. $\left(\tilde{\phi}_2^i\right)'\Big|_{0-} = M_{2\beta}^{-1}\left(\phi_\beta^i\right)'\Big|_{0-} = 1$ and

$$\left(\tilde{\phi}_1^i\right)'\Big|_{0-} = M_{1\beta}^{-1}\left(\phi_\beta^i\right)'\Big|_{0-} = 0 \tag{S14a}$$

$$\left(\tilde{\phi}_3^j\right)'\Big|_{0+} = N_{3\beta}^{-1}\left(\phi_\beta^j\right)'\Big|_{0+} = 0 \tag{S14b}$$

Case III: Source coming from the right through channel 3, i.e. $\left(\tilde{\phi}_3^j\right)'\big|_{0+} = N_{3\beta}^{-1}\left(\phi_\beta^j\right)'\big|_{0+} = 1$ and

$$\left(\tilde{\phi}_1^i\right)'\big|_{0-} = M_{1\beta}^{-1}\left(\phi_\beta^i\right)'\big|_{0-} = 0 \tag{S15a}$$

$$\left(\tilde{\phi}_2^i\right)'\big|_{0-} = M_{2\beta}^{-1}\left(\phi_\beta^i\right)'\big|_{0-} = 0 \tag{S15b}$$

## SIII. CURRENT AND NOISE AT D

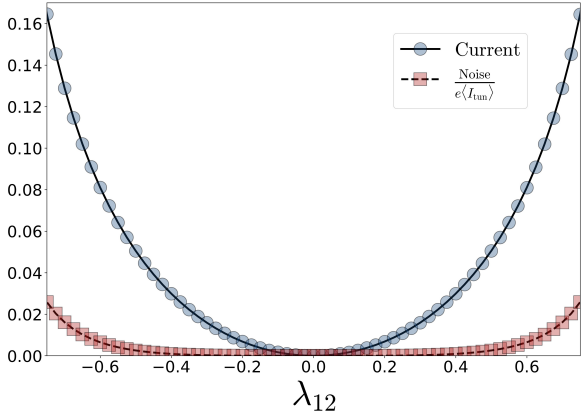

FIG. S2. Dc current and noise as a function of $\lambda_{12}$ for $\lambda_{23} = 0.5$ for velocities $v_1 = 1.0$, $v_2 = 1.1$ when both contacts are coupled.

First let us consider the case when the $C_1$ and $C_2$ are both coupled. Because they are perfect contacts [1] they absorb all the currents (charge or neutral) that reach region II and IV respectively. We will calculate the total fraction of the current injected at the source S that reaches the drain D. Different paths between S and D can be classified by the number of reflections at the boundaries $B_2$ and $B_3$ (these two numbers have to be equal). The greater the number of reflections, the longer the time to reach the drain. Therefore, the total current fraction at D, as a function of time is,

$$r_D(t) = r_{13}^{B_3} r_{31}^{B_2} \sum_{n=0}^{\infty} \Delta_n \delta(t - t_n). \tag{S16}$$

Here $\Delta_n = \left(r_{32}^{B_2} r_{23}^{B_3}\right)^n$ and

$$t_n = t_0 + n\Delta T, \tag{S17}$$

with $t_0$ being the shortest time to reach the drain and $\Delta T$ being the total time for one set of reflections at $B_2$ and $B_3$. Taking the Fourier transform, the current fraction at D as a function of $\omega$ is

$$r_D(\omega) = \frac{r_{13}^{B_3} r_{31}^{B_2}}{1 - r_{32}^{B_2} r_{23}^{B_3} e^{i\omega \Delta T}} \tag{S18}$$

We can now easily calculate the fraction of average tunneling current reaching the drain by summing over the all the charge packets reaching the drain as the zero frequency limit of this expression.

$$f = r_{13}^{B_3} r_{31}^{B_2} \left[1 + r_{32}^{B_2} r_{23}^{B_3} + \left(r_{32}^{B_2} r_{23}^{B_3}\right)^2 + \left(r_{32}^{B_2} r_{23}^{B_3}\right)^3 + \ldots\right]$$

$$= \frac{r_{13}^{B_3} r_{31}^{B_2}}{1 - r_{32}^{B_2} r_{23}^{B_3}} \tag{S19}$$

Next we compute the zero temperature noise following Ref. 2 for this problem.

$$L(t,t') = \langle I_D(t)I_D(t')\rangle + \langle I_D(t')I_D(t)\rangle - 2\langle I_D(t)\rangle\langle I_D(t')\rangle. \tag{S20}$$

Using $I_D(t) = r_D(t)I_1(t)$ we can write,

$$L(t, t+\tau) = r_D(t)r_D(t+\tau)S(\tau), \tag{S21}$$

with $S(\tau)$ being the conventional partitioning noise [2, 3]. Averaging over $t$ we get the noise

$$N_D(\tau) = h(\tau)S(\tau) \tag{S22}$$

$$\text{where } h(\tau) = \lim_{T\to\infty} \frac{1}{T}\int_0^T r_D(t)r_D(t+\tau)dt \tag{S23}$$

$$\text{and } S(\tau) = \frac{e^2\Gamma^2}{2\pi^2 a^2}\cos(V_0\tau)\frac{2\left(1 - \frac{v_1^2\tau^2}{a^2}\right)}{\left(1 + \frac{v_1^2\tau^2}{a^2}\right)^2}. \tag{S24}$$

Here the average tunneling current is $\langle I_{\text{tun}}\rangle = \frac{e\Gamma^2 V_0}{2\pi^2 v_1^2}$, where $\Gamma$ is the tunneling amplitude from the source into the probe charge chiral, $V_0$ is the potential difference between the source and the probe charge chiral, and $a$ is the UV cutoff ($a \to 0$ being the UV limit) [2]. Note that the dimension of $N_D(\tau)$ is the same as that of $S(\tau)$. Therefore, $N_D(\tau)$ has the dimensions of current squared and $N_D(\omega)/\langle I_{\text{tun}}\rangle$ has the dimensions of charge. Now, using Eq. S16 we can simplify $h(\tau)$ as,

$$h(\tau) = \left(r_{13}^{B_3}r_{31}^{B_2}\right)^2 \sum_{n,n'=0}^{\infty} \Delta_n \Delta_{n'} \delta\left(\tau + (t_n - t_{n'})\right). \tag{S25}$$

Thus the noise at zero frequency will be

$$N_D(\omega = 0) = \left(r_{13}^{B_3}r_{31}^{B_2}\right)^2 \sum_{n,n'=0}^{\infty} \Delta_n \Delta_{n'} S(t_{n'} - t_n) \tag{S26}$$

Now, using the expression for $S(\tau)$ from Eq. S24 in the limit $V_0 \to 0$ (small voltage difference) and $a \to 0$ we can easily find,

$$N_D(\omega = 0) = \frac{e\left(r_{13}^{B_3}r_{31}^{B_2}\right)^2}{1 - \left(r_{32}^{B_3}r_{23}^{B_3}\right)^2}\langle I_{\text{tun}}\rangle. \tag{S27}$$

Thus we can plot the noise to $e\langle I_{\text{tun}}\rangle$ ratio as a function of the interaction strength. We show one illustrative case in Fig. S2. When both contacts are coupled, noise is present at D but weak.

Next let us consider the case where the $C_1$ is coupled but $C_2$ is decoupled. Previously, when both contacts were coupled, all signals transmitted at $B_3$ were absorbed at $C_2$. Now, with $C_2$ decoupled, there will be multiple reflections in region IV. However, since there is no tunneling in any region, the $U(1)$ "charge" of each mode will be conserved. Thus the total reflected "charge" of the neutral (including all possible multiple reflections) from region IV to region III will vanish. Thus the dc current at the drain D will be zero.

## SIV. AN EXAMPLE WITH A CHARGED MODE IN THE TEST SYSTEM

As we understand that in the absence of the contacts $C_1$ & $C_2$ there will be no current at the drain D due to the charge conservation. This also guarantees that the total current at D will be zero if we remove either of the contacts. However, we can again calculate the total current at D when both the contacts are coupled. The first packet that boundary $B_3$ via mode 1 will reflect back to mode 2 of region III with a fraction $r_{12}^{B_3}$. The part that transmits to region IV will get absorbed by $C_2$. The packet that reflected into the mode 2 travels to boundary $B_2$ and one part reflects to mode 1 going towards the drain D with weight $r_{21}^{B_2}$, one part reflects to mode 3 travelling to boundary $B_3$ with weight $r_{23}^{B_2}$, and a part transmits to region II and gets absorbed by $C_1$. The part that travelled to boundary $B_3$ will reflect back to mode 2 with weight $r_{32}^{B_3}$ and the same process as the previous step starts. Thus the fraction of total current reaching the drain will be a series with multiple reflection in region III

$$f = r_{12}^{B_3} r_{21}^{B_2} \sum_{n=0}^{\infty} \left(r_{23}^{B_2} r_{32}^{B_3}\right)^n = \frac{r_{12}^{B_3} r_{21}^{B_2}}{1 - r_{23}^{B_2} r_{32}^{B_3}}. \quad (S28)$$

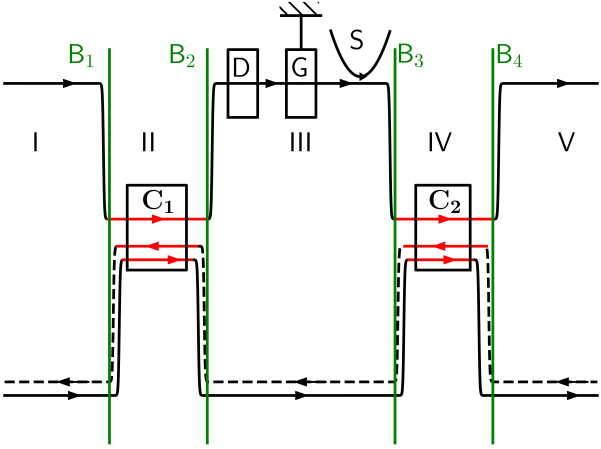

FIG. S3. A modified picture of the problem where the bottom region also has a right moving charged mode and a left moving neutral mode [4].

[1] C. Spånslätt, Y. Gefen, I. V. Gornyi, and D. G. Polyakov, Phys. Rev. B **104**, 115416 (2021), URL https://link.aps.org/doi/10.1103/PhysRevB.104.115416.
[2] T. Martin, Proceedings of the Les Houches Summer School, Session LXXXI (2005).
[3] C. d. C. Chamon, D. E. Freed, and X. G. Wen, Phys. Rev. B **51**, 2363 (1995), URL https://link.aps.org/doi/10.1103/PhysRevB.51.2363.
[4] C. L. Kane, M. P. A. Fisher, and J. Polchinski, Phys. Rev. Lett. **72**, 4129 (1994), URL https://link.aps.org/doi/10.1103/PhysRevLett.72.4129.



\end{document}